# Three levels of understanding physical relativity: Galileo's relativity, Up-to-date Galileo's relativity and Einstein's relativity: A historical survey.

**Bernhard Rothenstein**
Politehnica University of Timisoara, Physics Dept., Timisoara, Romania
E-mail: bernhard_rothenstein@yahoo.com

**Corina Nafornita**
Politehnica University of Timisoara, Communications Dept., Timisoara, Romania

Abstract
*We present a way of teaching Einstein's special relativity. It starts with Galileo's relativity, the learners know from previous lectures. The lecture underlines that we can have three transformation equations for the space-time coordinates of the same event, which lead to absolute clock readings, time intervals and lengths (Galileo's relativity), to absolute clock readings but to relative time intervals and lengths (up-to-date Galileo transformations) and to relative clock readings time intervals and lengths.*

## 1. First level of understanding physical relativity: absolute clock reading, time intervals and lengths

The student who starts learning Einstein's special relativity knows physics at a Newton-Galileo level, i.e. he is convinced that:

**A.** The true laws of physics are the same in all inertial reference frames and so it is impossible to distinguish between the state of motion and the state of rest of a laboratory by experiments we perform confined in it.
**B.** Relative motion does not desynchronize clocks once synchronized (brought to display the same time when they are instantly located at the same point in space) i.e. time and time intervals are absolute.
**C.** Measuring the length of the same rod observers in relative uniform motion obtain the same value i.e. lengths (space distances between two events) are absolute.
**D.** Distances measured perpendicular to the direction of relative motion have the same magnitude for all inertial observers in relative motion.
**E.** The coordinates of the same event, when measured from two inertial reference frames in relative motion transform in accordance with the following equations

$$x = x' + \mathrm{v}t' \quad (1)$$
$$y = y' \quad (2)$$
$$t = t' \quad (3)$$

where $(x, y, t)$ and $(x', y', t')$ are the space-time coordinates of the same event in the $K(XOY)$ and in the $K'(X'O'Y')$ reference frames respectively. The two frames are in the standard arrangement, v representing the velocity of $K'(X'O'Y')$ relative to K(XOY) in the positive direction of the overlapped OX(O'X') axes. Equations (1), (2) and (3) lead to the addition law for the OX(O'X') components of the velocity of the same particle

$$u_x = u'_x + vt' \qquad (4)$$

and to

$$u_y = u'_y \qquad (5)$$

for the OY(O'Y') components. We remind that an event represents a physical occurrence that could take place at a given point in space at a given time. The coordinates of the point where the event takes place represent its space coordinates whereas its time coordinate equates the reading of the clock located where the event takes place. We use for defining an event the notation $E(x, y, t)$ in K(XOY) and $E'(x', y', t')$ in K'(X'O'Y') respectively.

The art of the instructor who just starts to teach special relativity is to convince the learners that statements B, C and E hold only in the case when the involved velocities are small when we compare them with the velocity at which light propagates through empty space. However, he underlines that physicists hardly believe in statement A, which was not yet infirmed.

**F.** The student also knows that in all the experiments performed in a given inertial laboratory (Michelson, Foucault) we measure the two way velocity of light $c$. Doing that, we measure the time interval between the emission and the reception of a light signal that reflects itself on a mirror, located at a known distance from the source of light. Galileo describes such an experiment but with a negative result because the clocks he used where not able to measure very short time intervals.

**G.** There is no experimental evidence for the fact that light propagates forward at the same velocity $c_f$ at which it propagates backward $c_b$, $c_f$ and $c_b$ representing the one-way velocity of light.

**H.** Inertial observers in relative motion, equipped each with a co-moving light source obtain measuring the two-way velocity of light emitted by that source, the same value $c$ as a direct consequence of statement A.

## 2. The second level of understanding relativity: Absolute clock readings, relative time intervals and lengths

At that new level of understanding relativity we consider that the propagation of light is isotropic ($c_f = c_b$). In order to underline the important part played by clocks (devices able to generate a periodic phenomenon characterized by a strictly constant period) we consider that at each point of the OX axis of the K(XOY) reference frame we find a clock $C_i(x_i, y_i = 0)$ and let $C_0(x = 0, y = 0)$ be a clock located at the origin O of the K(XOY) reference frame. Let also be $C'_0(x'_i = 0, y'_i = 0)$ a clock located at the origin O'($x' = 0, y' = 0$) a clock located at the origin O' of the K'(X'O'Y') reference frame and commoving with it. All the clocks mentioned so far are identical. At $t = t' = 0$ when the origins of the two frames and implicitly the clocks $C_0$ and $C'_0$ are instantly located at the same point in space we synchronize the two clocks by simple comparison fixing them to display the same zero time.

Let $L(x = 0, y = 0)$ be a source of light located at $O(0,0)$ and at rest relative to $K(XOY)$. When clock $C_0$ reads $t_e$ source $L$ emits a light signal in the positive direction of the OX axis. Let $t_r$ be the reading of clock $C(x_i, y_i = 0)$ when the light signal arrives at its location. If

$$t_r = t_e + \frac{x_i}{c} \tag{6}$$

we say that we have synchronized the two clocks in accordance with a procedure proposed by Einstein. Under such conditions, all the clocks $C_i$ display the same running time. The moving clock $C'_0$ arrives successively in front of different $C_i$ clocks. Let $t'_{e,1}$ be the reading of clock $C'_0$ when it is instantly located in front of clock $C_i(x_i, 0)$ that reads $t_r$. A fundamental assumption in Galileo's relativity (statement B) is that the readings of the two clocks are equal to each other i.e.

$$t_r = t'_r. \tag{7}$$

Consider now that we attach a mirror $M'$ to clock $C'_0$. Source $L'$ emits a light signal when clock $C_0^0$ reads $t_e = 0$. The mirror located in front of source $L$ reflects it instantly back. At a time $t_e$, source $L$ emits a second light signal. When it arrives at the location of the mirror clock $C'_0$ reads $t'_{e,1}$, whereas a clock $C(x,0)$ in front of which it is instantly located reads $t_{e,1}$. In accordance with statement B we have $t_{e,1} = t'_{e,1}$. The obvious equation

$$v t_{e,1} = c(t_{e,1} - t_e) \tag{8}$$

equates the distance traveled by the mirror with the distance traveled by the second light signal and leads to the following relationship between the two clock readings

$$t_{e,1} = \frac{t_e}{1 - \dfrac{v}{c}}. \tag{9}$$

The reflected light signal arrives back at the location of clock $C_0^0$ when it reads $t_r$. It is obvious that now

$$v t_{e,1} = c(t_r - t_{e,1}) \tag{10}$$

from where we obtain

$$t_r = t_{e,1}\left(1+\frac{v}{c}\right). \qquad (11)$$

We illustrate the experiment described above using a classical space-time diagram (Figure 1).

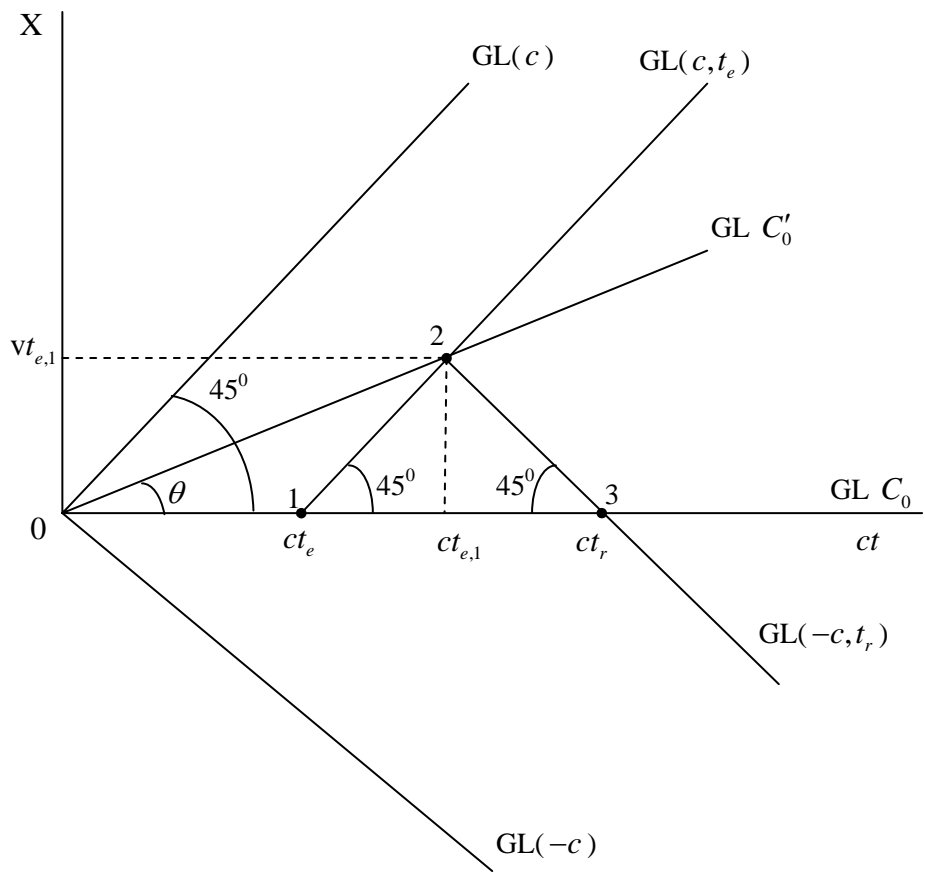

*Figure 1: We present an experiment, which enables us to find out the location and the reading of a moving clock using a classical space-time diagram. We use a classical space-time diagram that enables us to represent on it the geometric locus of the successive positions occupied by the moving clock and by the light signals propagating in the positive or in the negative direction of the OX axis.*

Instructors use such diagrams in teaching classical kinematics in order to define the instantaneous position of a particle at a given point of the OX axis ($x$) at a given time t. We use it to illustrate the geometric locus of the successive positions occupied by a moving particle, as well. Such a diagram presents two perpendicular axes on which we measure the instantaneous space coordinate ($x$) of a particle (space axis) and the product between the velocity of light through empty space c and the time $t$ when the particle occupies the mentioned position (time axis) respectively.
The following equation

$$x = \frac{v}{c}(ct) \tag{12}$$

describes the motion of the mirror and of the clock $C_0'$, attached to it. Theirs successive positions on the space-time diagram are located on the straight line GL $C_0'$, which makes with the time axis of the diagram an angle $\theta$ given by

$$\operatorname{tg} \theta = \frac{v}{c}. \tag{13}$$

The successive positions occupied by the light signals emitted at $t=0$ and $t_e$ respectively are located on the geometric loci $GL(c)$ and $GL(c,t_e)$ respectively, which make an angle $\theta_c$ with the positive direction of the time axis given by

$$\operatorname{tg} \theta_c = \frac{c}{c} = 1 \tag{14}$$

Light signals emitted in the negative direction of the OX axis appear on the space-time diagram as geometric loci that make an angle $\theta_{-c} = -45^0$ with its time axis ($GL(-c)$ and $GL(-c,t_r)$).

We consider now that we find the mirror $M(x,0)$ somewhere on the OX axis when a clock attached to it reads $t'$ whereas a clock $C(x,0)$ in front of which it is instantly located reads $t$. Source $L$ emits a light signal when clock $C_0^0$ reads $t_e$. It arrives at the location of clock $C(x,0)$ when it reads $t$. The mirror reflects it instantly back and it arrives at the location of clock $C_0^0$ when it reads $t_r$. From the obvious equations

$$x = c(t - t_e) \tag{15}$$

and

$$x = c(t_r - t) \tag{16}$$

we obtain

$$x = \frac{c}{2}(t_r - t_e) \tag{17}$$

$$t = \frac{1}{2}(t_r - t_e). \tag{18}$$

The important result is that using a single clock $C_0^0$ an observer located in front of the source $L$ could detect the position of the mirror $M(x,0)$ at a time $t$, simply measuring the times $t_e$ and $t_r$ when the source emitted the light signal and when it returned back to it. We illustrate the experiment in Figure 2.

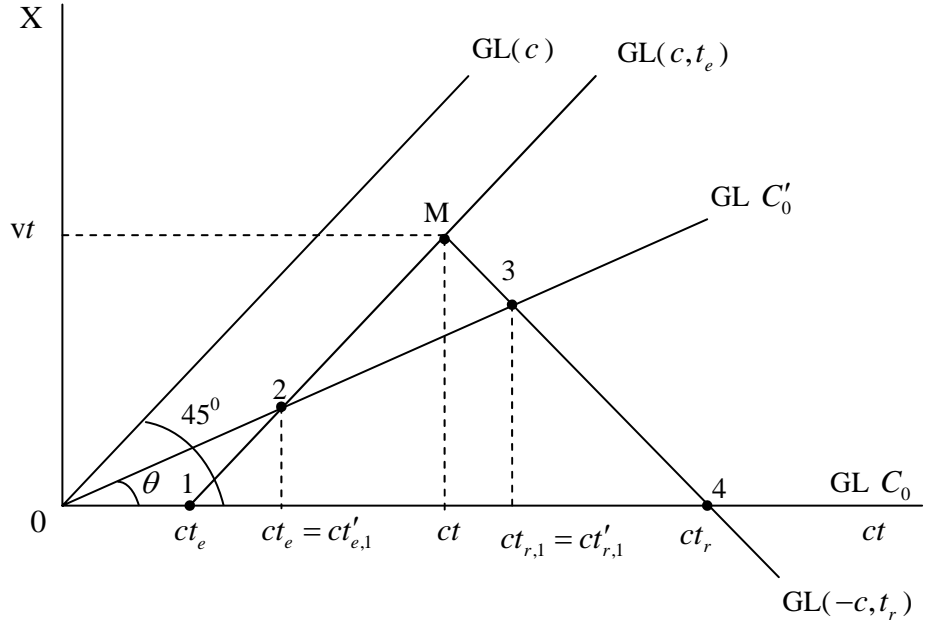

*Figure 2:* *We present a classical space-time diagram used in order to illustrate how two inertial observers in relative motion could determine the reading and the location of a clock located somewhere on the overlapped OX(O'X') axes.*

We find at the origin of the $K'(X'O'Y')$ reference frame a source of light $L'$. The classical space-time diagram we show in Figure 2 tells us that in order to illuminate the mirror at the same position on the OX axis, source $L'$ should emit the light signal at a time $t'_{e,1} = t_{e,1}$ receiving it back at a time $t'_{r,1} = t_{r,1}$. In accordance with statement A we should have

$$x' = \frac{c}{2}(t'_{r,1} - t'_{e,1}) \tag{19}$$

$$t' = \frac{1}{2}(t'_{e,1} + t'_{r,1}) \tag{20}$$

The results obtained performing the experiment illustrated in Figure 1 enable us to consider that (equations (9) and (11))

$$t_{e,1} = t'_{e,1} = \frac{t_e}{1-\frac{v}{c}} \tag{21}$$

and that

$$t_{r,1} = t'_{r,1} = \frac{t_e}{1+\frac{v}{c}}. \tag{22}$$

Expressing the right hand side of (17) and (18) as a function of times measured in $K'(X'O'Y')$ and taking into account (19) and (20) we obtain successively

$$x = \frac{c}{2}(t_r - t_e) = \frac{c}{2}(t'_{r,1} - t'_{e,1}) + \frac{v}{2}(t'_{r,1} + t'_{e,1}) = x' + vt' \tag{23}$$

$$t = \frac{1}{2}(t_r + t_e) = \frac{1}{2}(t'_{r,1} + t'_{e,1}) + \frac{v}{2c}(t'_{r,1} - t'_{e,1}) = t' + \frac{v}{c^2}x'. \tag{24}$$

We call equations (23) and (24) up-to-date Galileo transformations and they lead to the following addition law of velocities orientated in the positive direction of the overlapped axes $OX(O'X')$

$$u_x = \frac{x}{t} = \frac{x' + vt'}{t' + \frac{v}{c^2}} = \frac{u'_x + v}{1 + \frac{v}{c^2}u'_x} \tag{25}$$

in which we recognize Einstein's addition law of the corresponding components. If $u'_x = c$ then (25) leads to $u_x = c$ as well in accordance with Einstein's postulate: Measuring the velocity at which light propagates in empty space, observers from different inertial reference frames obtain for it the same value $c$. For the $OY(O'Y')$ component of the velocity we obtain

$$u_y = \frac{y}{t} = \frac{y'}{t' + \frac{v}{c^2}x'} = \frac{u'_y}{1 + \frac{v}{c^2}u'_x}. \tag{26}$$

As we see, there are some problems with (25) and (26) because they do not fulfill the fundamental condition imposed by Einstein

$$\sqrt{u_x^2 + u_y^2} = \sqrt{u'^2_x + u'^2_y} \tag{27}$$

in the case when in both reference frame observers measure the velocity of a light signal.

**3. What is wrong with the up-to-date Galileo transformations? Third level of understanding relativity: relative clock readings, time intervals and lengths**

We concentrate on equation (24). For $x' = 0$ it leads to the result

$$t = t' = 0 \tag{28}$$

i.e. the readings of clocks $C'_0$ and $C(x = vt, y = 0)$ when they are instantly located at the same point in space are equal to each other, a hypothesis with which we started the derivation of the up-to-date Galileo transformations. We say that two events are simultaneous in a given reference frame when they take place there at the same time. As an example, events $E_1'(x_1', y_1', t')$ and $E_2'(x_2', y_2', t')$ are simultaneous in the $K'(X'O'Y')$ reference frame and take place at different two points in space. The up-to-date Galileo transformations lead to the conclusion that the two events are no longer simultaneous if we detect them from the $K(XOY)$ reference frame. Event $E_1'$ and $E_2'$ detected from $K(XOY)$ have the following time coordinates

$$t_1 = t' + \frac{v}{c^2} x_1' \tag{29}$$

$$t_2 = t' + \frac{v}{c^2} x_2'. \tag{30}$$

As a consequence, the time separation between the two events is in $K(XOY)$

$$\Delta t = t_2 - t_1 = \frac{v}{c^2}(x_2' - x_1') \tag{31}$$

being proportional with the space separation between the two events in $K'(X'O'Y')$. That is a paradoxical situation because absolute time readings and relative simultaneity are irreconcilable physical conditions.

In order to remove the paradox we look for a correct relationship between the readings of two clocks in relative motion when they are instantly located at the same point in space, in our case clocks $C(x, y)$ and $C'(x' = 0, y' = d)$. We consider the synchronization of clocks $C_0'(0,0)$ and $C'(0,d)$ in the $K'(X'O'Y')$ reference frame (Figure 3a) and in the $K(XOY)$ reference frame (Figure 3b).

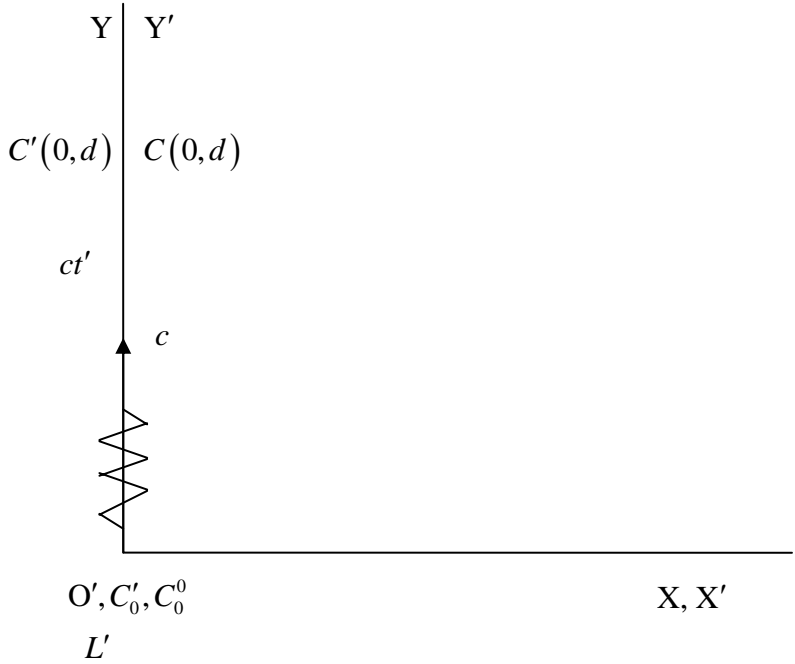

**Figure 3a:** *Synchronization of two clocks at rest in the $K'(X'O'Y')$ reference frame.*

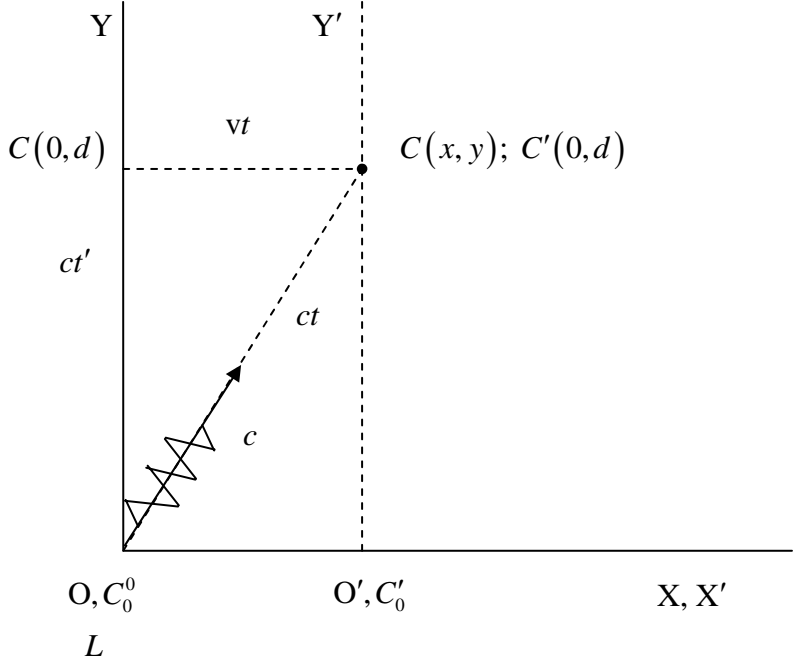

**Figure 3b:** *Synchronization of the two clocks at rest in $K'(X'O'Y')$ as detected from the $K(XOY)$ reference frame.*

When clock $C'_0(0,0)$ reads $t'=0$, a source of light $L'(0,0)$ located at $O'$ and at rest in $K'(X'O'Y')$ emits a light signal in the positive direction of the $O'Y'$ axis. Arriving at the location of clock $C'(0,d)$, stopped and reading $\dfrac{d}{c}$, the light signal starts it. After that procedure, clocks $C'_0$ and $C'(0,d)$ display the same running time and we say that we synchronized them in accordance with a synchronization procedure proposed by Einstein.

When detected from the $K(XOY)$ reference frame, clock $C'(0,d)$ is instantly located in front of clock $C(x,y)$ synchronized with clock $C_0^0(0,0)$ in accordance with the synchronization procedure proposed by Einstein. When clock $C(x,y)$ reads $t$, clock $C'(0,d)$ has advanced with $vt$ in the positive direction of the OX axis whereas the light signal, which performed the synchronization, traveled a distance $ct$. Because of the invariance of distances measured perpendicular to the direction of relative motion, Pythagoras' theorem leads to

$$c^2t^2 = v^2t^2 + c^2t'^2 \qquad (32)$$

from where

$$t = \frac{t'}{\sqrt{1-\dfrac{v^2}{c^2}}} \qquad (33)$$

We underline that (33) holds only in the case when one of the involved clocks has a $x' = 0$ space coordinate. In order to lead to the result imposed by (31) equation (24) should read

$$t = \frac{t' + \frac{v}{c^2} x'}{\sqrt{1 - \frac{v^2}{c^2}}} \qquad (34)$$

In order to lead to (25) equation (23) should read

$$x = \frac{x' + vt'}{\sqrt{1 - \frac{v^2}{c^2}}}. \qquad (35)$$

With (2) and (34) equation (26) takes the correct shape

$$u_y = \frac{y}{t} = \frac{u'_y \sqrt{1 - \frac{v^2}{c^2}}}{1 + \frac{vu'_x}{c^2}} \qquad (36)$$

We have now all the elements for deriving all the relativistic effects, encountered in relativistic kinematics.

## 4. Conclusions

As a moral of our historical survey we can consider that some times, a physicist finds himself in a similar situation to that of a tailor who has to fix the coat, which he has tailored (the theory) in such a way that it matches the client's body (mother nature). Commonly speaking, he cuts there and ads elsewhere. Rigorously, he looks for a tailoring method which enables him to consider all the peculiarities of the body and which may ensure the success from the first "fitting". Good experimental results and experimentally proved postulates ensure the success from the first fitting.
Telling the learners that we have used in our derivations the "police radar" and the "radar detection procedure" we could make them more receptive.